\documentclass[useAMS,usenatbib,a4paper,preprint]{mn2e}
\usepackage{graphicx}
\usepackage{amsmath}
\usepackage{txfonts}
\usepackage{mathrsfs}
\usepackage{multicol}
\usepackage{color}

\def\aap{A\&A}
\def\apjl{ApJ}

\def\apjs{ApJS}

\def\apjs{ApJ Supplements}

\def\apjl{ApJ Letters}

\def\aap{A\&A}

\newcommand{\be}{\begin{equation}}
\newcommand{\ee}{\end{equation}}
\newcommand{\bea}{\begin{eqnarray}}
\newcommand{\eea}{\end{eqnarray}}

\title[Model selection using time-delay lenses]{Model selection using time-delay lenses}
\author[Fulvio Melia]{Fulvio Melia,$^1$\thanks{John Woodruff Simpson
Fellow. E-mail: fmelia@email.arizona.edu} Jun-Jie Wei,$^{2,3}$ and Xue-Feng Wu$^{2,3}$ \\
$^1$Department of Physics, The Applied Math Program, and Department of Astronomy,
The University of Arizona, AZ 85721, USA\\
$^2$Purple Mountain Observatory, Chinese Academy of Sciences, Nanjing 210023, China\\
$^3$School of Astronomy and Space Sciences, University of Science and Technology of China, Hefei 230026, China}

\begin{document}

\date{}

\pagerange{\pageref{firstpage}--\pageref{lastpage}} \pubyear{2021}

\maketitle

\label{firstpage}

\begin{abstract}
The sample of time-delay gravitational lenses appropriate for studying the
geometry of the Universe continues to grow as dedicated campaigns, such as
the Dark Energy Survey, the VST ATLAS survey, and the Large Synoptic Survey
Telescope, complete their census of high-redshift sources. This catalog
now includes hundreds of strong lensing systems, at least 31 of which have
reasonably accurate time delay measurements. In this paper, we use
them to compare the predictions of two competing
Friedmann-Lema\^itre-Robertson-Walker models: flat $\Lambda$CDM, characterized
by two adjustable parameters ($H_0$ and $\Omega_{\rm m}$), and the
$R_{\rm h}=ct$ universe (with $H_0$ as the single free variable). Over the
past decade, the latter has accounted for the data better than the standard
model, most recently the emergence of well-formed galaxies discovered by
{\it JWST} at cosmic dawn. Here we show that the current sample of
time-delay lenses favours $R_{\rm h}=ct$ with a likelihood of $\sim84\%$
versus $\sim16\%$ for the standard model. This level of accuracy will greatly
improve as the ongoing surveys uncover many thousands additional lens systems
over the next several years.
\end{abstract}


\section{Introduction}\label{introduction}
The use of time-delay gravitational lenses to measure the expansion
of the Universe has been considered for almost sixty years, starting with the
original proposal by \cite{Refsdal:1964}. Light rays originating from a
single quasar source travel through different gravitational potentials on
either side of a foreground lensing galaxy, so their deflection angles and
travel times probe the model-dependent angular diameter distance from the
source to the lens and the lens to the observer \citep{Petters:2001}.

Of the hundreds of strong lensing systems already discovered, a small fraction
of them have exhibited measurable time delays between the various images. In
this paper, we assemble 31 of these with reasonably accurate data
one can use for model selection. About half of the lenses have been added to the
catalog only recently from the COSmological MOnitoring of GRAvItational Lenses
survey (COSMOGRAIL; \citealt{Eigenbrod:2005}), the Dark Energy Survey (DES;
\citealt{Banerji:2008,Schneider:2014}, the Large Synoptic
Survey Telescope (LSST; \citealt{Tyson:2002})
project, and the VST ATLAS campaign \citep{Koposov:2014}.

It is already well recognized that these strong lensing sources can yield
useful constraint on the parameters in the standard model
\citep{Paraficz:2009,Suyu:2013}. For example, they may eventually
help to resolve the current disagreement between the Hubble constant measured
at low and high redshifts \citep{Rathna:2015,Denzel:2021}. But their usefulness
extends well beyond a single application to $\Lambda$CDM. They may also
help with model selection via the comparative testing of quite diverse
cosmologies, as we have already attempted to do using a much smaller sample
(of only 12 time-delay lens systems) available back then \citep{Wei:2014}.

Our analysis is motivated in part by the growing tension seen between
the predictions of the standard model and the actual observations. For example,
there now exists a significant disparity between the value of the Hubble
constant, $H_0$, inferred locally and its measurement based on the statistical
analysis of anisotropies in the cosmic microwave background (CMB; \citealt{Riess:2022}).

The Hubble constant ($=67.4\pm0.5$ km $\rm s^{-1}$ $\rm Mpc^{-1}$) measured
by {\it Planck} \citep{PlanckVI:2020} is in $5.0\sigma$ tension with that measured
using Type Ia supernovae, calibrated via the Cepheid distance ladder
($H_0=73.04\pm1.04$ km $\rm s^{-1}$ $\rm Mpc^{-1}$). Other early-Universe probes,
such as clustering and weak lensing, yield results similar to the CMB
\citep{Abbott:2018a}. Measurements based on the red giant branch settle on
an intermediate value of $H_0$ \citep{Freedman:2019,Freedman:2020}. Given that
the errors associated with these measurements are probably realistic \citep{Verde:2019},
the implied $5.0\sigma$ disparity in the expansion of the Universe at low and high
redshifts seems to refute the standard model's predicted evolution of the Universe.

But the problems with $\Lambda$CDM extend well beyond this well-known inconsistency
with $H_0$. A recently published review \citep{Melia:2023} highlights at least
eight independent areas where the standard model is in significant
conflict with either the observations or fundamental physical principles. A brief
survey of these inconsistencies includes the initial entropy problem, an unknown
classicalization process that must have converted seed quantum fluctuations into the
perturbations responsible for the growth of structure, the incorrect timeline implied
by the early appearance of supermassive black holes and well-formed galaxies, and the
incorrect prediction of light-element abundances (notably the $^7$Li anomaly) during
big bang nucleosynthesis in this model.

It is therefore instructive to continue comparing the predictions of $\Lambda$CDM
with those of another Friedmann-Lema\^itre-Robertson-Walker cosmology, known as
the $R_{\rm h}=ct$ universe \citep{Melia:2007,MeliaShevchuk:2012}. Over the past 15
years, this model has been tested using over 27 different kinds of cosmological
data, at both high and low redshifts, and has been favoured over the standard
model by the various model selection criteria. A recent compilation of these
results may be found in Table~2 of \cite{Melia:2018e}. Its theoretical basis
and a more complete examination of its consistency with the data may be found
in \cite{Melia:2020}

In this paper, we update our earlier analysis of time-delay lenses with the improved
statistics offered by a sample almost three times larger than before, and examine whether
our previous conclusion favouring $R_{\rm h}=ct$ over $\Lambda$CDM changes as
the sample size grows.

In \S~\ref{lensing}, we briefly summarize the key steps required to use
time-delay lenses in cosmological testing, and then apply them to our sample
of 31 systems in \S~\ref{delay}, where we report the outcome of the one-on-one
comparison between $\Lambda$CDM and $R_{\rm h}=ct$. We end with our
conclusion in \S~\ref{conclusion}.

\begin{table*}\label{tdelay}
\caption{Time Delay Lenses}
\begin{tabular}{lllccccccl}
\hline\hline
{\rm System}&$\quad z_l$&$\quad z_s$&$\theta_A$&$\theta_B$&$\Delta t=t_A-t_B$&${\mathcal{R}}_{\rm obs}$
(with $\eta=0.29$)& ${\mathcal{R}}_{R_{\rm h}=ct}$&${\mathcal{R}}_{\Lambda{\rm CDM}}$&{\rm Refs.} \\
&&&({\rm arcsec})&({\rm arcsec})&(days)&(Gpc)&(Gpc)&(Gpc)& \\
\hline
{\rm FBQ}0951+2635   &  0.26  &  1.246  &  $0.886\pm0.004$ &  $0.228\pm0.008$ &  $-16.0\pm2.0$    &
$1.238\pm0.391$  &   0.947    &  0.927  & 6 \\
{\rm SDSS J}1442+4055 & 0.284 & 2.593 & $1.385\pm0.049$ & $0.771\pm0.028$ & $-25.0\pm1.5$  &
$1.051\pm0.331$  &   0.893    &  0.889  & 36, 37 \\
{\rm RX J}1131-1231  &  0.295 &  0.657  & $1.898\pm0.015$  & $1.922\pm0.023$  &  $1.6\pm0.7$      &
$0.963\pm1.215$  &   1.509    &  1.469  & 21, 39 \\
{\rm PG}1115+0.80 & 0.311     & 1.722 & $1.12\pm0.014$  &  $0.95\pm0.016$  &  $-8.3\pm1.5$            &
$1.286\pm0.467$  &   1.045    &  1.036  & 32, 33 \\
{\rm QJ}0158-4325    &  0.317 &  1.29   & $0.814\pm0.038$  & $0.41\pm0.02$  &  $-22.7\pm3.6$      &
$2.491\pm0.884$  &   1.156    &  1.141  & 21, 40 \\
{\rm Q}0957+561      &  0.36  &  1.413  &  $5.220\pm0.006$ &  $1.036\pm0.11$  &  $-417.09\pm0.07$ &
$0.837\pm0.243$  &   1.282    &  1.272  & 6, 13, 14\;\; \\
{\rm HS}0818+1227    &  0.39  &  3.113  & $2.219\pm0.009$  & $0.615\pm0.003$  &  $-153.8\pm13.9$  &
$1.740\pm0.529$  &   1.140    &  1.152  & 21 \\
{\rm SDSS J}0924+0219 & 0.393 &  1.523  & $0.878\pm0.021$  & $0.977\pm0.072$  &  $2.4\pm3.8$      &
$0.671\pm1.203$  &   1.368    &  1.363  & 21, 41, 6 \\
{\rm SDSS J}1620+1203 & 0.398 &  1.158  & $2.277\pm0.037$  & $0.494\pm0.01$  &  $-171.5\pm8.7$    &
$1.775\pm0.526$  &   1.567    &  1.552  & 21, 42 \\
{\rm DES}2325-5229   &  0.400 &  2.74   & $1.23\pm0.05$ & $1.79\pm0.05$ &  $43.8\pm4.3$           &
$1.322\pm0.439$  &   1.191    &  1.201  & 24, 46 \\
{\rm HE}0047-1756    &  0.407 &  1.678  & $0.894\pm0.043$ & $0.553\pm0.027$ &  $-10.4\pm3.5$      &
$1.071\pm0.508$  &   1.371    &  1.370  & 21, 24, 38 \\
{\rm B}1600+434      &  0.414 &  1.589  &  $1.14\pm0.075$  &  $0.25\pm0.074$  &  $-51.0\pm4.0$    &
$2.084\pm 0.692$  &  1.422    &  1.421  & 4, 5, 20 \\
{\rm J}1001+5027     &  0.415 &  1.838  &  $1.98\pm0.08$ &  $0.93\pm0.19$ &  $-119.3\pm3.3$       &
$1.972\pm0.651$   &  1.357    &  1.360  & 30, 31 \\
{\rm SDSS J}1335+0118 & 0.44  &  1.570  & $1.079\pm0.031$ & $0.489\pm0.014$ &  $-56.0\pm5.9$      &
$3.004\pm0.954$   &  1.523    &  1.524  & 21, 48 \\
{\rm WG}0214-2105    &  0.45  &  3.24   & $1.018\pm0.225$ & $1.246\pm0.275$ &  $7.5\pm2.8$        &
$0.716\pm1.193$   &  1.273    &  1.296  & 24, 47 \\
{\rm HE}0435-1223    &  0.454 &  1.693  & $1.298\pm0.008$ & $1.168\pm0.014$ &  $-9.0\pm0.8$       &
$1.380\pm0.451$   &  1.527    &  1.533  & 21, 6 \\
{\rm Q}0142-100      &  0.491 &  2.73   &  $1.855\pm0.002$ &  $0.383\pm0.005$ &  $-97\pm15.8$     &
$1.411\pm0.469$   &  1.419    &  1.445  & 6, 12, 21 \\
{\rm SDSS J}1650+4251 & 0.577 & 1.547  &  $0.872\pm0.027$ &  $0.357\pm0.042$ &  $-49.5\pm1.9$    &
$3.544\pm1.083$   &  2.079    &  2.102  & 6, 18 \\
{\rm DES J}0408-5354 &  0.597 &  2.375  & $3.626\pm0.342$ & $3.053\pm0.361$ &  $112.1\pm2.1$      &
$1.311\pm1.199$   &  1.758    &  1.802  & 26, 43 \\
{\rm HE}2149-2745    &  0.603 &  2.033  &  $1.354\pm0.008$  &  $0.344\pm0.012$ &  $-72.6\pm17.0$  &
$1.887\pm0.704$   &  1.890    &  1.929  & 6, 9, 21, 49 \\
{\rm SDSS J}1339+1310 & 0.609 & 2.231 & $0.580\pm0.041$ & $1.234\pm0.106$ & $47.0\pm5.5$          &
$1.760\pm0.677$   &  1.835    &  1.879  & 34, 35 \\
{\rm SDSS J}0832+0404 & 0.659 &  1.116  & $1.56\pm0.024$ & $0.435\pm0.008$ &  $-125.3\pm18.2$     &
$2.405\pm0.784$   &  3.469    &  3.494  & 21, 45 \\
{\rm B}0218+357      &  0.685 &  0.944  &  $0.057\pm0.004$  &  $0.280\pm0.008$ &  $+11.3\pm0.2$   &
$6.378\pm1.892$   &  5.312    &  5.313  & 1, 2, 3, 19 \\
{\rm Q}1355-2257     &  0.702 &  1.370  & $0.959\pm0.081$ & $0.267\pm0.023$ &  $-81.5\pm11.6$     &
$4.034\pm1.500$   &  3.005    &  3.052  & 21, 6 \\
{\rm SBS}1520+530    &  0.717 &  1.855  &  $1.207\pm0.004$  &  $0.386\pm0.008$ &  $-130.0\pm3.0$  &
$4.137\pm1.204$   &  2.397    &  2.460  & 6, 16 \\
{\rm HE}1104-1805    &  0.729 &  2.319  &  $1.099\pm0.004$  &  $2.095\pm0.008$ &  $152.2\pm3.0$   &
$1.978\pm0.575$   &  2.150    &  2.223  & 2, 7, 8\;\; \\
{\rm SDSS J}1515+1511 & 0.742 &  2.054  & $1.676\pm0.104$ & $0.313\pm0.019$ &   $-210.2\pm5.6$    &
$3.181\pm1.013$   &  2.338    &  2.411  & 21, 44 \\
{\rm SDSS J}1206+4332 & 0.748 &  1.789  &  $1.870\pm0.088$  &  $1.278\pm0.097$ &  $-111\pm3$      &
$2.435\pm0.891$   &  2.588    &  2.659  & 17, 23 \\
{\rm SBS}0909+532     & 0.83  &  1.377  &  $0.415\pm0.126$  &  $0.756\pm0.152$ &  $+50.0\pm3.0$   &
$4.890\pm3.415$   &  4.034    &  4.127  & 6, 15, 22 \\
{\rm PKS}1830-211     & 0.89  &  2.507  &  $0.67\pm0.08$    &  $0.32\pm0.08$   &  $-26.0\pm5.0$   &
$2.837\pm1.386$   &  2.523    &  2.640  & 10, 11 \\
{\rm WFI J}2026-4536  & 1.04  &  2.23   & $0.673\pm0.156$ & $0.801\pm0.184$ &   $18.7\pm4.8$      &
$3.472\pm6.795$   &  3.290    &  3.464  & 21, 6 \\
\hline\hline
\end{tabular}
{References: {(1) \cite{Carilli:1993}; (2) \cite{Lehar:2000};
(3) \cite{Wucknitz:2004}; (4) \cite{Jackson:1995}; (5) \cite{Dai:2005};
(6) \cite{Kochanek:2008}; (7) \cite{Wisotzki:1993}; (8) \cite{Poindexter:2007};
(9) \cite{Burud:2002}; (10) \cite{Lovell:1998}; (11) \cite{Meylan:2005};
(12) \cite{Koptelova:2012}; (13) \cite{Falco:1997}; (14) \cite{Colley:2003};
(15) \cite{Dai:2009}; (16) \cite{Auger:2008}; (17) \cite{Paraficz:2009};
(18) \cite{Vuissoz:2007}; (19) \cite{Biggs:2018}; (20) \cite{Burud:2000};
(21) \cite{Millon:2020a}; (22) \cite{Hainline:2013}; (23) \cite{Eulaers:2013};
(24) \cite{Millon:2020b}; (25) \cite{Koopmans:2003}; (26) \cite{Courbin:2018};
(27) \cite{Eulaers:2011}; (28) \cite{Fohlmeister:2008}; (29) \cite{Bonvin:2019};
(30) \cite{Rathna:2013}; (31) \cite{Oguri:2005}; (32) \cite{Bonvin:2018};
(33) \cite{Morgan:2008a}; (34) \cite{Goicoechea:2016}; (35) \cite{Inada:2009};
(36) \cite{Shalyapin:2019}; (37) \cite{Sergeyev:2016}; (38) \cite{Wisotzki:2004};
(39) \cite{Sluse:2007}; (40) \cite{Morgan:2008b}; (41) \cite{Inada:2003};
(42) \cite{Kayo:2010}; (43) \cite{Agnello:2017}; (44) \cite{Inada:2014};
(45) \cite{Oguri:2008}; (46) \cite{Ostrovski:2017}; (47) \cite{Spiniello:2019};
(48) \cite{Oguri:2004}; (49) \cite{Rathna:2015}.}}
\end{table*}

\section{Strong Lensing}\label{lensing}
Time-delay gravitational lenses constitute a powerful probe of the underlying
cosmology, but are limited by our imprecise knowledge of the central lens mass
distribution and other possible perturbing lenses along the line-of-sight. One
can simplify the procedure by assembling a subsample of homogeneous systems for
which one may reasonably assume the same lens model to adequately represent
the mass distribution in every case \citep{Oguri:2004}. Of course, we
already know the selected sample is not perfectly homogeneous from the few
lens systems that have been modeled independently, suggesting at least some
variation in the lens structure. Taking a statistical approach, as we do here,
however, can still be useful if there is a way to quantify the non-negligible
dispersion arising from the poorly known systematics, which include the
non-uniformity of the lens itself. We discuss how this will be done in
\S~\ref{delay}. We shall find via the use of Equation~(\ref{eq:sigmasys})
that the dispersion due to this variation of the lens mass distribution
appears to be smaller than $\sim 29\%$ of the measured time-delay distance
(Eq.~\ref{eq:tddistane}). While not ideal, this relatively modest
inhomogeneity does present us with a workable approach we can use
for model selection purposes (see \S~\ref{model.selection}).

Let the source be located at angle $\vec{\beta}$. The time delay, $\Delta t_i$,
for an image $i$ at angular position $\vec{\theta}_i$ is due to the difference
in path length along the deflected and undeflected null geodesics, as well as
the gravitational time dilation incurred by the rays traversing the gravitational
potential, $\Psi(\vec{\theta}_i)$, of the lens:
\begin{equation}
\Delta t_i = {1+z_l\over c}{D_A(0,z_s)\,D_A(0,z_l)\over D_A(z_l,z_s)}\left[{1\over 2}
(\vec{\theta}_i-\vec{\beta})^2-\Psi(\vec{\theta}_i)\right]\label{eq:Deltat}
\end{equation}
\noindent\citep{Blandford:1986}. In this expression, $z_l$ and $z_s$ are the lens and source
redshifts, respectively, and $D_A(z_1,z_2)$ is the angular diameter distance between
$z_1$ and $z_2$. Thus, when the lens geometry $\vec{\theta}_i-\vec{\beta}$ and potential
$\Psi$ are known, the time delay yields the so-called time-delay distance,
\begin{equation}
{\mathcal{R}}\equiv {D_A(0,z_s)\,D_A(0,z_l)\over D_A(z_l,z_s)}\;,\label{eq:tddistane}
\end{equation}
which clearly depends on the cosmological model.

One may find a replacement for $\Psi(\vec{\theta}_i)$ by noting that the mass
distribution in lens spiral and elliptical galaxies can be represented quite
well by a power-law density profile \citep{Rusin:2003}, for which
\begin{equation}
\Psi(\vec{\theta})={b^2\over 3-n}\left({\theta\over b}\right)^{3-n}\;,\label{eq:Psi}
\end{equation}
in terms of the deflection scale, $b$, and index, $n$. The special case with $n=2$
is the singular isothermal sphere (SIS), with
\begin{equation}
b=4\pi{D_A(z_l,z_s)\sigma^2_\star\over D_A(0,z_s)}\;,\label{eq:b}
\end{equation}
where $\sigma_\star$ is the velocity dispersion of the lensing galaxy.
Observations of the galaxy density distributions seem to suggest that $n$ is
often close to the isothermal value, so the SIS is both convenient and
accurate for the majority of lens galaxies \citep{Koopmans:2009}.
The time delay between two images at $\vec{\theta_A}$ and $\vec{\theta_B}$ in such
a lens system is given as
\begin{equation}
\Delta t=t_A-t_B={1+z_l\over 2c}\left(\theta_B^2-\theta_A^2\right)
{\mathcal{R}}(z_l,z_s)\;,\label{eq:Dt}
\end{equation}
when the velocity dispersion of an SIS is assumed in these expressions.
With this approach, the `observed' time-delay distance is therefore
inferred from the expression
\begin{equation}
{\mathcal{R}}_{\rm obs}(z_l,z_s)={2c\over 1+z_l}{\Delta t\over
\left(\theta_B^2-\theta_A^2\right)}\;.\label{eq:Robs}
\end{equation}

Of course, the actual velocity dispersion $\sigma_\star$ may not match the
analytical value exactly, but these do appear to be very close in the majority
of lensing systems for which $\sigma_\star$ has been measured \citep{Treu:2006}.
Nevertheless, the number of time-delay lenses for which $\sigma_\star$ is
known observationally is too small to make this datum useful, so all of the
analysis we carry out in this paper is based solely on the use of
Equation~(\ref{eq:Dt}), though we allow for possible variations of the
velocity dispersion from system to system via the introduction of a
systematic error defined in Equation~(\ref{eq:sigmasys}).

\section{Time-delay lenses}\label{delay}
\subsection{The Sample and Methodology}
Our sample of 31 time-delay gravitational lenses for which ${\mathcal{R}_{\rm obs}}$
has been `measured' with Equation~(\ref{eq:Dt}) is shown in Table~1.
All of these are gravitationally lensed quasars, which show enough variability
to permit an accurate measurement of a time-delay between the various images. Six of
them are quadrupole image systems (RX J1131-1231, PG1115+0.80, WG0214-2105, HE0435-1223,
DES J0408-5354, and WFI J2026-4536). The rest are two-image systems.

Columns~2--6 display the observational data, while column~7 lists the value of
${\mathcal{R}_{\rm obs}}$, together with its uncertainty $\sigma_{\mathcal{R}}$.
This dispersion is calculated from the statistical error using the propagation
equation
\begin{equation}
\sigma_{\rm stat}=\mathcal{R}_{\rm obs}\left[\left({\sigma_{\Delta t}\over \Delta t}\right)^{2}
+4\left({\theta_B\sigma_{\theta_B}\over \theta_B^2-\theta_A^2}\right)^2
+4\left({\theta_A\sigma_{\theta_A}\over \theta_B^2-\theta_A^2}\right)^2
\right]^{1/2}\;,\label{eq:staterror}
\end{equation}
and a second source of (systematic) error, $\sigma_{\rm sys}$, that takes
into account various effects giving rise to the observed scatter of individual
lenses from the assumed pure SIS profile. These include a possible rms deviation of
the lens velocity dispersion, a steepening of the mean mass density profile
compared to that of a pure SIS \citep{Koopmans:2009}, and the non-zero line-of-sight
contribution, which is non-zero on average.

\begin{figure}\label{fig1}
\center{\includegraphics[scale=0.55,angle=0]{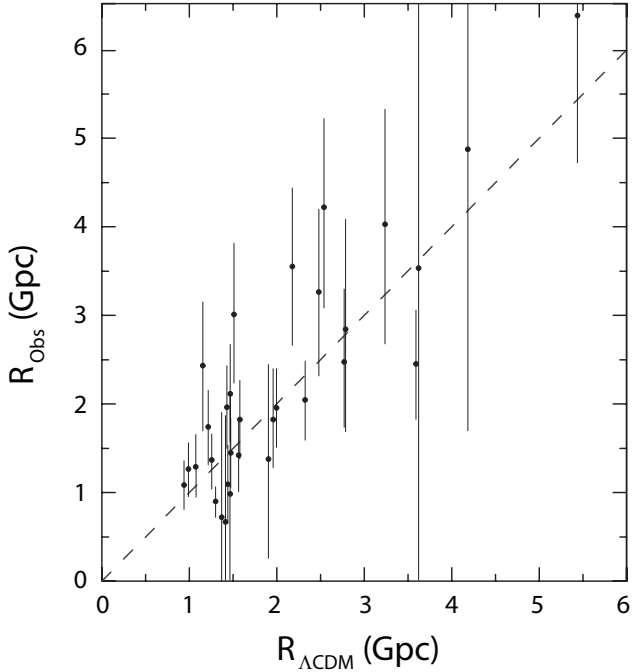}
\caption{Thirty one ${\mathcal{R}}$ measurements, with error bars, compared to the
predictions of $\Lambda$CDM, with optimized parameters: $H_{0}=85.6^{+7.0}_{-5.9}\;(1\sigma)$
km $\rm s^{-1}$ $\rm Mpc^{-1}$, $\Omega_{\rm m}=0.50^{+0.34}_{-0.34}\;(1\sigma)$ and
$\eta=0.29^{+0.06}_{-0.05}\;(1\sigma)$ (see Fig.~2). The dashed line indicates a perfect
match between theory and observation. The reduced $\chi^2$ for this fit is
$0.971$ ($28$ dof).}}
\end{figure}

According to \cite{Cao:2012}, $\sigma_{\rm sys}$ may be as large as $\sim 20\%$.
Since we don't know how large this error is a priori, we model it along with the
other variables in our maximum likelihood analysis and write it as
\begin{equation}
\sigma_{\rm sys}\equiv  \eta\,\mathcal{R}_{\rm obs}\;,\label{eq:sigmasys}
\end{equation}
in terms of the additional free parameter $\eta$. Its introduction is just a
convenient way to characterize this systematic uncertainty in terms of the
value of ${\mathcal{R}}_{\rm obs}$. In our maximum likelihood estimation,
this additional free parameter is optimized along with the other free parameters
individually for each model being tested. As we shall see below, however, $\eta$
appears to be quite independent of the cosmology itself, and turns out to have the
value $\eta=0.29^{+0.06}_{-0.05}\;(1\sigma)$ for the sample of 31 lens systems used
in this study, consistent with the earlier finding of \cite{Cao:2012}. The total
uncertainty $\sigma_{\mathcal{R}}$ in ${\mathcal{R}}_{\rm obs}$ is then found by adding
$\sigma_{\rm stat}$ and $\sigma_{\rm sys}$ in quadrature,
\begin{equation}
\sigma_{\mathcal{R}}^{2}=\sigma_{\rm stat}^{2}+\sigma_{\rm sys}^{2}\;,\label{eq:sigtot}
\end{equation}
and this is the error appearing in column 7 of Table~1.

Columns 8 and 9 in this table display the optimized model predictions for
$R_{\rm h}=ct$ and $\Lambda$CDM, respectively, which we now describe.
The angular-diameter distance in $\Lambda$CDM is a function of several
parameters, including $H_0$ and the mass fractions $\Omega_{\rm m}
\equiv \rho_{\rm m}/\rho_{\rm c}$, $\Omega_{\rm de}\equiv \rho_{\rm de}/
\rho_{\rm c}$, and $\Omega_{\rm r}\equiv \rho_{\rm r}/\rho_{\rm c}$,
for matter, dark energy and radiation, respectively, defined as ratios
of the critical density $\rho_{\rm c}\equiv 3c^2H_0^2/8\pi G$.
To streamline the optimization of the fit with this model, we adopt a
minimal number of unknown variables, so we assume that dark energy is
a cosmological constant ($\Omega_{\rm de}=\Omega_\Lambda$), and that
the spatial curvature constant is zero. In that case, since the contribution
from radiation in the redshift range covered in Table~1 is insignificant
compared to that of the others, we simply put $\Omega_{\rm m}+\Omega_\Lambda=1$.
Then, the angular-diameter distance between redshifts $z_1$ and $z_2$
($>z_1$) is given by the expression
\begin{equation}
D_A^{\;\Lambda{\rm CDM}}(z_1,z_2)={c\over H_0}{1\over (1+z_2)}\int_{z_1}^{z_2}
\frac{dz}{\left[\Omega_{\rm m}(1+z)^3+\Omega_\Lambda\right]^{1/2}}\;.\label{eq:dAL}
\end{equation}
With this application of the standard model, we thus have three free parameters
in total for the optimization: the cosmological parameters $H_0$ and $\Omega_{\rm m}$,
and the time-delay lens dispersion parameter $\eta$.

On the other hand, the angular-diameter distance in $R_{\rm h}=ct$
\citep{Melia:2007,MeliaShevchuk:2012,Melia:2020} depends only on $H_0$:
\begin{equation}
D_A^{\;R_{\rm h}=ct}(z_1,z_2)={c\over H_0}{1\over (1+z_2)}
\ln\left({1+z_2\over 1+z_1}\right)\;.\label{eq:dAR}
\end{equation}
To find the best fit to the data in Table~1, we therefore optimize two
free parameters for this model: $H_0$ and $\eta$.

For each model, we optimize the fit by maximizing the joint likelihood function
\begin{equation}
L(\sigma_{\mathcal{R}},\xi)\propto\prod_{i=1}
\frac{1}{\sigma_{\mathcal{R}_i}}\;
\exp\left[-\frac{\left(\mathcal{R}_{\rm\emph{i}}[\xi]-\mathcal{R}_{\rm obs,
\emph{i}}\right)^{2}} {2\sigma_{\mathcal{R}_i}^2} \right]\;,\label{eq:like}
\end{equation}
where ${\mathcal{R}}_{\rm\emph{i}}[\xi]$ is the theoretical prediction
of the time-delay distance between $z_{l,i}$ and $z_{s,i}$ for
either $\Lambda$CDM or $R_{\rm h}=ct$ and the model specific parameters $\xi$,
$\mathcal{R}_{\rm obs, \emph{i}}$ is the measured value, and $\sigma_{\mathcal{R}_i}$
is the dispersion of $\mathcal{R}_{\rm obs, \emph{i}}$ given in Equation~(\ref{eq:sigtot}).

\begin{figure}\label{fig2}
\center{\includegraphics[scale=0.45,angle=0]{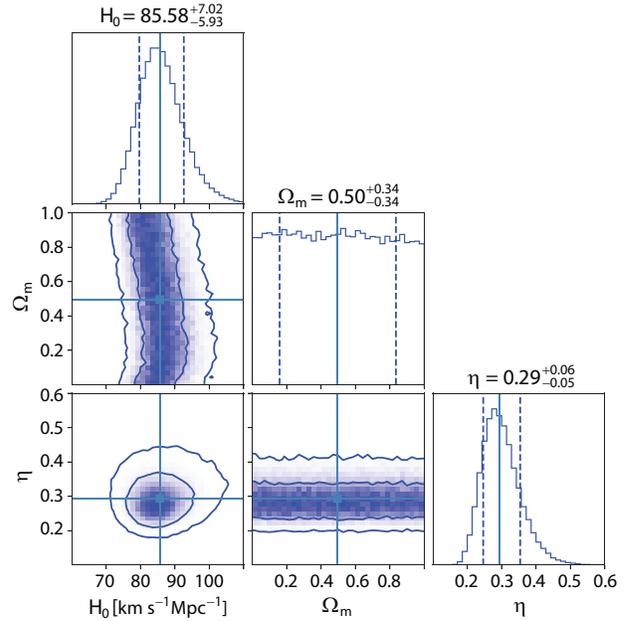}
\caption{1D probability distributions and 2D regions with $1\sigma$ and $2\sigma$ contours
for the best-fit $\Lambda$CDM model (see Fig.~1).}}
\end{figure}

We follow this procedure using the data in Table~1 with the restriction
$0.0\le\Omega_{\rm m}\le 1.0$, and find that $\Lambda$CDM fits the observed time-delay
distances with a maximum likelihood corresponding to the parameter values
$H_{0}=85.6^{+7.0}_{-5.9}$ ($1\sigma$) km $\rm s^{-1}$ $\rm Mpc^{-1}$,
$\Omega_{\rm m}=0.50^{+0.34}_{-0.34}$ ($1\sigma$) and $\eta=0.29^{+0.06}_{-0.05}$ ($1\sigma$).
The best fit with the $R_{\rm h}=ct$ universe is provided by the optimized parameters
$H_{0}=81.3_{-5.4}^{+6.5}$ ($1\sigma$) km $\rm s^{-1}$ $\rm Mpc^{-1}$ and
$\eta=0.29^{+0.06}_{-0.05}$ ($1\sigma$). The $1\sigma$ and $2\sigma$ confidence intervals
for these model are shown in Figures~2 and 4, and the entries in columns~8 and 9 of Table~1
correspond to these best-fit parameters.

These results confirm a feature of the fit with $\Lambda$CDM reported
on several previous occasions \citep{Coe:2009,Linder:2011,Treu:2016}. The principal
dependence of the time-delay distance (Eq.~\ref{eq:tddistane}) is on the Hubble
constant, $H_0$, to which it is inversely proportional. The impact of a
variation of $\Omega_{\rm m}$ on the ratio of angular-diameter distances in
this expression is much weaker. One therefore needs a much bigger sample
to constrain $\Omega_{\rm m}$ very tightly.

The observed values ${\mathcal{R}}_{\rm obs}$ are displayed in Figures~1 and 3,
in comparison with those predicted by the two models. In these plots, a perfect match
would correspond to the straight dashed line. The clustering of points towards small values
of ${\mathcal{R}}$ reflects the preponderance of lenses at relatively small redshifts.
Interestingly, the optimized value of $\eta$ is the same in both models, suggesting that
the systematics are independent of the cosmology and do indeed represent an intrinsic
dispersion of the lens potential away from a pure SIS. With its inclusion in the overall
error budget, we find that both models fit this sample of 31 time-delay lenses quite well.
The reduced $\chi^2$ is $0.971$ ($28$ dof) for $\Lambda$CDM in Figure~1, and a slightly
better $0.937$ ($29$ dof) for $R_{\rm h}=ct$ in Figure~3.

\subsection{Model Selection}\label{model.selection}
Since these models formulate the angular-diameter distance in Equations~(\ref{eq:dAL})
and (\ref{eq:dAR}) differently, however, without the same number of free parameters,
a comparison of the likelihoods to determine which of them is favoured by the data
needs to be based on model selection tools. To compare the evidence for and against
competing models, the use of information criteria has become quite common in cosmology
(see, e.g., \citealt{Takeuchi:2000,Liddle:2004,Liddle:2007,Tan:2012,Schwarz:1978}).
Such a tool may be viewed as an enhanced `goodness of fit' criterion, extending the
usual~$\chi^2$ diagnostic by including the number of free parameters in each model.
The information criteria favour models with fewer parameters over those with many,
as long as the latter do not provide a considerably better fit to the data.

\begin{figure}\label{fig3}
\center{\includegraphics[angle=0,scale=0.55]{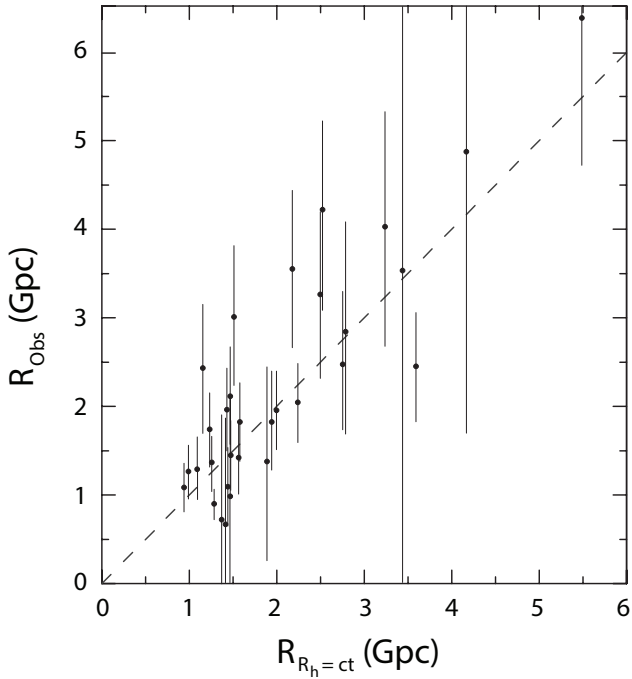}
\caption{Same as Fig.~1, except now for the $R_{\rm h}=ct$ universe,
with optimized parameters: $H_{0}=81.3^{+6.5}_{-5.4}\;(1\sigma)$
km $\rm s^{-1}$ $\rm Mpc^{-1}$ and $\eta=0.29^{+0.06}_{-0.05}\;(1\sigma)$
(see Fig.~4). The reduced $\chi^2$ for this fit is $0.937$ ($29$ dof).}}
\end{figure}

An information criterion provides the relative ranks of two or more competing models,
along with a numerical estimate of confidence that each is the best, analogous to
the likelihood (or posterior probability) in traditional statistical inference.
Information criteria are superior to the latter, however, in that they can be applied
to models that are not `nested,' providing a procedure for comparing candidates that
are not specializations of each other.

An information criterion can be applied after the following kind of regression is
performed. Suppose that corresponding to the values $z_1,\dots,z_N$ of a free
parameter the measured values are $h_1,\dots,h_N$ of a dependent one, with (known)
error bars $\pm\sigma_1,\dots,\pm\sigma_N$. These errors are assumed to be normally
distributed.  Now let a model~$\mathcal{M}$ predict values $\hat h_1,\dots,\hat h_N$,
inferred from a formula $\hat h_i=\hat h_i(\vec\beta)$ involving a parameter
vector~$\vec{\beta}$ comprising $f$~unknown variables, i.e., $\vec{\beta}=
(\beta_1,\dots,\beta_f)$. In other words, the data model~$\mathcal{M}$ is in
effect a statistical one, of the form
\begin{equation}
  h_i = \hat h_i({\vec\beta}) + \sigma_i Z_i\;,\label{eq:stat1}
\end{equation}
where $Z_1,\dots,Z_N$ are independent standard normal random variables.
Note that in the case of linear regression, $\hat h_i(\vec\beta)$~would be
$\sum_{j=1}^f X_{ij}\beta_j$ for known coefficients~$X_{ij}$. Typically,
$X_{ij}=\hat h^{(j)}(z_i)$ for known functions $\hat h^{(1)},\dots,\hat
h^{(f)}$ of~$z$.

\begin{figure}\label{fig4}
\center{\includegraphics[scale=0.63,angle=0]{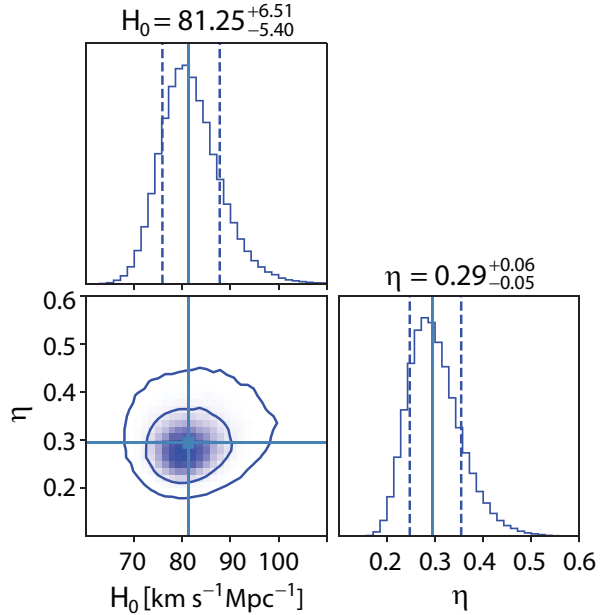}
\caption{Same as Fig.~2, except now for the $R_{\rm h}=ct$ universe, whose
best fit is shown in Fig.~3.}}
\end{figure}

Then for model~$\mathcal{M}$, the $\chi^2$ goodness of fit to the data is
expressible as
\begin{equation}
\chi^2 = \sum_{i=1}^N [h_i - \hat h_i(\vec\beta)]^2/\sigma_i^2\;,\label{eq:stat2}
\end{equation}
i.e., a (weighted) sum of squared errors, and the $\chi^2$ per degree of freedom,
\begin{equation}
\chi^2_{\rm{dof}} = \chi^2 / (N-f)\;.\label{eq:stat3}
\end{equation}
By necessity, we must also have $N>f$.  The parameters $(\beta_1,\dots,\beta_f)$ are
selected to minimize the~$\chi^2$, producing a best fit to the data.

For the data we use in this paper, the sample is large (i.e., $>20$), so the most
appropriate one to use is the Bayesian Information Criterion (BIC; \citealt{Schwarz:1978}),
which tests the statistical performance of the models.  It is defined as
\begin{equation}
{\rm BIC}\equiv -2\ln L+\left(\ln N\right)f\label{eq:BIC}
\end{equation}
where, as previously noted, $N$ is the number of data points (here $31$) and $f$
is the number of free parameters (three for $\Lambda$CDM and two for $R_{\rm h}=ct$).

The BIC is generally considered to be a large-sample ($N\gg 1$) approximation to the outcome
of a conventional Bayesian inference procedure for selecting the model preferred by the data.
Among the available choices being tested, the model with the lowest BIC score is the one
selected by this criterion. Among its many proponents, \cite{Liddle:2007} has
argued for the use of BIC in cosmological model selection, which has now been done to
compare several popular models against $\Lambda$CDM (see, e.g., \citealt{Shi:2012}).

Two statistical models of the data set $(h_1,\dots,h_N)$, such as a
`true' model $\mathcal{M}_*$ and a second model~$\mathcal{M}$, may be
viewed as probability density functions (PDF's) on~$\mathbb{R}^N$, say
$F_*(h_1,\dots,h_N)$ and $F(h_1,\dots,h_N)$, respectively.  In information
theory, the discrepancy of the PDF $F$ from the PDF~$F_*$---effectively,
a measure of distance---is given by the Kullback--Leibler formula
\begin{equation}
  D(\mathcal{M}_*\|\mathcal{M}) = \int_{\mathbb{R}^n}
{\rm d}h_1\dots {\rm d}h_n\, F_*(h)\,\ln\frac{F_*(h)}{F(h)}\,\ge\,0
\end{equation}
(using the notation that the argument $h$ stands for the entire
data set $[h_1,\dots,h_N]$).  The best model~$\mathcal{M}$ may be
selected from a set of candidate models by choosing the one with the
minimum $D(\mathcal{M}_*\|\mathcal{M})$.  The model $\mathcal{M}_*$~is
not known, but the $\mathcal{M}$ with parameters optimized by
minimizing~$\chi^2$, is special.  It turns
out that the BIC of the fitted model~$\mathcal{M}$ approximates
$2D(\mathcal{M}_*\|\mathcal{M})$, up~to an unimportant additive constant.
This is especially truen when $\mathcal{M}_*$ is a model of the same
type, with free variables $(\beta_1^*,\dots,\beta_f^*)$.

The quantity ${\rm BIC}/2$ is an unbiased estimator of the distance
$D(\mathcal{M}_*\|\mathcal{M})$. This statement is exact for linear
regression, and correct to leading order for non-linear regression.
The fitted model~$\mathcal{M}$ does depend on the data set, so
both $D(\mathcal{M}_*\|\mathcal{M})$ and $\textrm{BIC}/2$ are random
variables.  In the language of probability, a lack of bias
translates to them having the same expectation.

The extent to which the fitted information criteria represent an
\emph{accurate} estimate of $2D(\mathcal{M}_*\|\mathcal{M})$ has been
investigated theoretically (see, e.g., \citealt{Yanagihara:2005}).
Their variability has also been studied by repeatedly comparing
$\Lambda$CDM to other cosmological models \citep{Tan:2012}. In the
case of BIC, it is known that the magnitude of the difference
$\Delta=\allowbreak {\rm BIC}_2 -\nobreak {\rm BIC}_1$ provides
a numerical assessment of the evidence that model~1 is preferred
over model~2.  As a general rule, the evidence is considered to be
weak if $\Delta\la2$, mildly strong if $\Delta\approx3$ or~$4$,
and quite strong if $\Delta\ga5$.

One may thus rank the competing models quantitatively as follows. If
${\rm BIC}_\alpha$ characterizes model $\alpha$, the unnormalized confidence that this model
is correct is the `Bayes weight' $\exp(-{\rm BIC_\alpha}/2)$. The likelihood of model
$\alpha$ being correct relative to the others is then
\begin{equation}
P(\alpha)=\frac{\exp(-{\rm BIC_\alpha}/2)}{\sum_{i}\exp(-{{\rm BIC}_{i}}/2)}\;.\label{eq:Palpha}
\end{equation}
The sum in the denominator includes all of the candidates being tested simultaneously.

The outcome of this analysis shows that $R_{\rm h}=ct$ is preferred over $\Lambda$CDM with
a $-2\ln L=444.73$ versus $444.68$ (or equivalently, with a ${\rm BIC}=451.60$ versus $454.98$),
which translates into a probability of $\sim84\%$ versus $\sim16\%$ of being the correct model. To complete the discussion, we also considered how well the time-delay lens data are fit by {\it Planck} $\Lambda$CDM,
i.e., the version of the standard model with its parameters fixed at the values optimized by
{\it Planck} \citep{PlanckVI:2020}: $H_0=67.4$ km $\rm s^{-1}$ $\rm Mpc^{-1}$ and
$\Omega_{\rm m}=0.315$. Not surprisingly, a head-to-head comparison between this
model and $R_{\rm h}=ct$ reveals a much more slanted result, with the latter being favoured
by these data with a probability of $\sim 99.3\%$ versus only $\sim 0.7\%$.

\section{Conclusion}\label{conclusion}
Our use of a much larger time-delay lens sample ($31$) compared to the set of 12 systems we
used in our first attempt to carry out this kind of analysis \citep{Wei:2014}, has refined
the parameter search and reduced their $1\sigma$ errors, but the optimized values themselves have
changed very little. For example, the updated value of $H_0$ in the
context of $\Lambda$CDM is now $85.6^{+7.0}_{-5.9}\;(1\sigma)$ km $\rm s^{-1}$ $\rm Mpc^{-1}$
compared with $=87^{+17}_{-16}\;(1\sigma)$ km $\rm s^{-1}$ $\rm Mpc^{-1}$. But the intrinsic
dispersion (as measured by the parameter $\eta$) persists, as indicated by its current value,
$\eta=0.29^{+0.06}_{-0.05}\;(1\sigma)$, versus $\eta=0.29^{+0.15}_{-0.09}$ ($1\sigma$).
As noted earlier, this suggests that the poorly known systematics are probably due solely
to deviations of the lens potential from a pure SIS, as opposed to any cosmological
influence.

Overall, our analysis of the time-delay distances using a larger sample size has affirmed
our earlier conclusion that $R_{\rm h}=ct$ is favoured by these data over the current
standard model. This catalog of lens systems is expected to grow considerably over the
coming years, improving the statistics even further and motivating a more realistic
handling of the lens mass distribution in order to mitigate the impact of $\sigma_{\rm sys}$
well below the current limitations.

In our previous work \citep{Wei:2014}, we estimated that if the real
cosmology is $\Lambda$CDM, one would need a sample of $\sim 150$ time-delay lenses
to rule out $R_{\rm h} = ct$ at a confidence level of $\sim 99.7\%$, while $\sim 1000$
lenses would be required to rule out $\Lambda$CDM if the actual universe were instead
$R_{\rm h} = ct$. This difference is due to the greater flexibility afforded the
standard model by its greater number of free parameters.

Looking forward, the prospects of attaining a sample of this size look quite promising.
Several ongoing and future survey programs include a search for strong lensing systems
in their purview. According to some estimates, the Large-aperture Synoptic Survey Telescope
(LSST) will observe some 8000 lensed quasars during its 10-year lifetime, about 3000
of which will have well-measured time-delays \citep{Oguri:2010,Collett:2015}. A subsample
useful for a detailed analysis based on the modelling of individual lenses would require an
accurate characterization of the mass distribution of the lens galaxy, auxiliary data including
high-resolution imaging, and measurements of the stellar velocity distribution. Such
requirements would reduce the sample of strong gravitational lenses with accurate
time-delay measurements to $\sim 100$. With the statistical approach we have used
in this paper, however, in contrast to individual modelling, the latter requirements may
generally be waived, allowing us to expect a usable sample approaching 1000 or more
new (LSST) lenses for future work.

But LSST is not the only survey mission promising to increase the sample size. {\it Euclid}
may provide a comparable number \citep{Jee:2016,Laureijs:2011}, and the {\it DESI} Legacy Imaging
Survey has already started reporting the discovery of over 1000 new strong lensing
candidates (most of which are galaxy-galaxy pairs, rather than quasar-galaxy
systems). Future work is expected to uncover many thousands more \citep{Huang:2022}.
The Square Kilometer Array (SKA) will in addition be able to compile up to
$\sim 10^5$ lensing systems with $\sim 100\%$ reliability and known source
redshifts \citep{Laureijs:2011}. Several hundred new candidates have also been
reported by the (HOLISMOKES) Hyper Suprime-Cam (HSC) survey, with thousands
more expected with continued imaging \citep{Canameras:2021}.

Our expectation is that the requisite number of strong lensing systems with accurate
time-delay measurements to rule out $R_{\rm h}=ct$ if the real cosmology is
$\Lambda$CDM should be attainable in just a few years. Compiling a sample of
$\sim 1000$ such systems to rule out $\Lambda$CDM if the real cosmology is
instead $R_{\rm h}=ct$ will take longer, but is nevertheless quite feasible
during this decade with the combined efforts of the many ongoing and future
survey missions.

\section*{Acknowledgments}
We are very grateful to the anonymous referee for their excellent review of
this paper and for making several suggestions to improve its presentation. This work
is partially supported by the National Key Research and Development Program of China
(2022SKA0130100), the National Natural Science Foundation of China (grant Nos. 11725314
and 12041306), the Key Research Program of Frontier Sciences (grant No. ZDBS-LY-7014)
of Chinese Academy of Sciences, and the Natural Science Foundation of Jiangsu Province
(grant No. BK20221562).
\vfill\newpage

\section*{DATA AVAILABILITY STATEMENT}
No new data were generated or analysed in support of this research.

\bibliographystyle{mn2e}
\bibliography{ms}

\label{lastpage}
\end{document}